\documentclass[symmetry,article,accept,pdftex,oneauthor]{Definitions/mdpi}

\graphicspath{{figs/}}
\firstpage{1} 
\pubvolume{1}
\issuenum{1}
\articlenumber{0}
\pubyear{2025}
\copyrightyear{2025}
\externaleditor{ Ignatios Antoniadis}
\datereceived{28 November 2024 } 
\daterevised{27 December 2024 } % Comment out if no revised date
\dateaccepted{7 January 2025 } 
\datepublished{ } 
\hreflink{https://doi.org/} % If needed use \linebreak
\Title{Gauge Covariance of the Gap Equation: From the Rainbow Truncation to Gauge Symmetry Constraints}

\TitleCitation{Gauge Covariance of the Gap Equation: From the Rainbow Truncation to Gauge Symmetry Constraints}

\Author{Bruno El-Bennich$^{1,2}$\orcidA{}}

\AuthorNames{Bruno El-Bennich}

\AuthorCitation{El-Bennich, B.}
\address{%
$^{1}$ \   Departamento de F\'isica, Universidade Federal de S\~ao Paulo, \\ \quad  Rua S\~ao Nicolau 210, Diadema, 09913-030 S\~ao Paulo, Brazil;  bennich@unifesp.br \\ 
$^{2}$ \   Instituto de F\'isica Te\'orica, Universidade Estadual Paulista\\ \quad  Rua Dr.~Bento Teobaldo Ferraz 271, S\~ao Paulo, 01140-070 S\~ao Paulo, Brazil}

\abstract{The gauge covariance of the quark gap equation is compared for the case of three different quark--gluon vertices: the bare vertex, a Ball--Chiu-like vertex  constrained by the corresponding Slavnov--Taylor identity, and a full vertex including the transverse components derived from transverse Slavnov--Taylor identities.  The covariance properties are verified with the chiral quark condensate and the pion decay constant in the chiral limit. }


\begin{document}

%%%%%%%%%%%%%%%%%%%%%%%%%%%%%%%%%%%%%%%%%%%%%%%%%%%%%%%%%%%%%%%%%%%%%%%%%%%%%%%%%%%%%%%%%
%%%%%%%%%%%%%%%%%%%%%%%%%%%%%%%%%%%%%%%%%%%%%%%%%%%%%%%%%%%%%%%%%%%%%%%%%%%%%%%%%%%%%%%%%

\section{Symmetries: When to Break, When to Preserve}

One of the great achievements of hadron physics in the past decades has been the demonstration that \emph{dynamical chiral symmetry breaking\/} is overwhelmingly 
responsible for the masses of nucleons and atoms~\cite{Bashir:2012fs,Yang:2018nqn}. Indeed, while the Brout-Englert-Higgs mechanism is the source of explicit chiral 
symmetry breaking, expressed by current-quark mass terms in the Standard Model Lagrangian, the symmetry breaking due to gluon dynamics is incomparably more 
efficient in generating the mass scales we observe in nuclear physics. This is possible because in describing strong interactions with a relativistic quantum field theory, 
namely Quantum Chromodynamics (QCD), we also deal with the implicit Einstein formula $E=mc^2$.  In other words, most of the visible mass  we observe is due to 
radiation energy. It seems as though \emph{Nature} is telling us in a twisted manner that perfect symmetry is not always what she aims at.

On the other hand, gauge symmetry and its preservation have been the \emph{leitmotif} in developing relativistic quantum field theories during the past  century and 
are still guiding physicists in the pursuit of Standard Model extensions. This is a natural demand for any theory, as one can redefine the fields and particles by 
a gauge transformation in such a way that the physical laws remain the same. Simply put, gauge symmetries are nothing else but redundant degrees of freedom or 
an over-complete description of a physical system. One may do whatever one wishes with the superfluous degrees of freedom, as the physics of a given system remains
the same and any measurable quantities cannot depend on how we choose them. This is commonly called gauge invariance and is also intimately related to the 
renormalizability of a quantum field theory. For instance, the Ward-Green-Takahashi identity (WGTI)~\cite{Ward:1950xp,Green:1953te,Takahashi:1957xn}
that describes current conservation in Quantum Electrodynamics (QED) implies that the wave function renormalization of the electron and its vertex renormalization 
constant are equal. This identity is crucial to the cancellation of the ultraviolet divergences that occur in loop calculations to all orders in perturbation theory. 

Now, while we insist on gauge symmetry to be preserved at all times, we still have to deal with the freedom to redefine fermion and gauge fields. The consequence is
that their equations of motion, or their Green functions, are altered. This brings gauge covariance into play, which means that the Green functions are not gauge invariant 
though they obey well-defined transformations with respect to the local gauge group. How these Green functions specifically transform under gauge variation is described 
by the Landau-Khalatnikov-Fradkin transformations (LKFT)~\cite{Landau:1955zz,Fradkin:1955jr} within the class of linear covariant gauges. If these transformations are 
correctly applied, physical observables derived from Green functions in any given gauge should be \emph{identical\/}. This is exactly what we call gauge invariance. 
We recall that the Nielsen identities also provide a means to relate variations of Green functions under a gauge parameter change~\cite{Nielsen:1975fs,DeMeerleer:2019kmh}.

In practice, the application of the LFKT is rather cumbersome, as they are written in coordinate space. One must first obtain a nonperturbative solution for a Green 
function in a given gauge, for instance the quark propagator in Landau gauge, and then Fourier transform this solution to coordinate space. After this one applies 
the LFKT and Fourier transforms this gauge-transformed propagator back to the momentum space for any arbitrary covariant gauge. This is an arduous and nontrivial 
task and the procedure was shown to be feasible with analytical expressions in leading truncation schemes in QED~\cite{Bashir:2004yt}. In an SU($N$) gauge 
field theory and for covariant $R_\xi$ gauges, the transformation law for the quark propagator was derived with a perturbative expansion of the SU($N$) gauge transformation 
of the quark field to $\mathcal{O}(g_s^6)$ in the corresponding two-point Green function~\cite{Aslam:2015nia}. Another derivation of the LFKT in non-Abelian theories 
was given in Ref.~\cite{DeMeerleer:2018txc}. 

On the other hand, since the LFKT are nonperturbative in nature, the initial propagator should also be the solution of the nonperturbative gap equation. The latter is 
described by a Dyson-Schwinger equation (DSE) which involves, besides the boson and fermion propagators, the fully-dressed fermion-boson vertex. The LFKT themselves 
do not tell us anything about the general form of this vertex, but we can make use of the WGTI identity in QED or of the Slavnov-Taylor identity (STI) in QCD~\cite{Slavnov:1972fg,
Taylor:1971ff} to constrain at least its non-transverse parts, and also of transverse WGTIs~\cite{Kondo:1996xn,He:2000we,He:2006my,He:2007zza,Pennington:2005mw} 
and STIs~\cite{He:2009sj} to deduce the transverse vertices~\cite{Albino:2018ncl,Albino:2021rvj,El-Bennich:2022obe}. 

Recently, an alternative path to the LFKT was taken comparing the solutions of the quark DSE in different covariant $R_\xi$ gauges~\cite{Lessa:2022wqc}. 
The gauge dependence  in this gap equation is twofold, for the dependence on $\xi$ enters directly via the gluon dressing function in $R_\xi$ gauges, obtained in lattice 
QCD simulations~\cite{Bicudo:2015rma} up to $\xi=0.5$, and the strong coupling $\alpha_s^\xi$. Indirectly, the dressed quark-gluon vertex also contributes to this 
dependence and this feature will be exploited in the present study. The numerical data of the $\xi$-dependent gluon propagators was fitted with a Pad\'e approximant 
which revealed that the fit parameters exhibit a nearly linear dependence on the gauge parameter. This feature was taken advantage of and the gluon propagator was 
then extrapolated to Feynman gauge. Employing the full quark-gluon vertex derived from Slavnov-Taylor identities and gauge covariance in Ref.~\cite{Albino:2021rvj}, 
the quark's mass and wave renormalization functions were computed in the range $\xi \in [0,1]$. The quark condensate derived therefrom exhibits a modest 
dependence on the gauge parameter within the error estimates of the lattice predictions for the gluon, in agreement with a prediction of the LKFT in 
QCD~\cite{Aslam:2015nia}.

We here take the opportunity to extend the study of Ref.~\cite{Lessa:2022wqc} by including simpler truncations of the DSE for comparison and to highlight their effect 
on the gauge covariance of the gap equation. As we shall see, the functional $\xi$-dependence of the quark-mass function is diametrically opposed when the DSE is 
solved in either the rainbow truncation or with the full vertex structure. Moreover, the quark condensate and pion decay constant, the gauge-parameter dependence of
the latter presented here for the first time, decrease as functions of $\xi$ in this leading truncation scheme in disagreement with the gauge-invariance predictions of 
the LFKT. To a certain extent this is expected, as the rainbow-ladder truncation badly violates the WGTI and STI and is therefore commonly employed in Landau gauge 
to ``minimize'' the error. Nonetheless, we here explicitly demonstrate the failure of a simpler truncation to satisfy gauge covariance.

%%%%%%%%%%%%%%%%%%%%%%%%%%%%%%%%%%%%%%%%%%%%%%%%%%%%%%%%%%%%%%%%%%%%%%%%%%%%%%%%%%%%%%%%%
%%%%%%%%%%%%%%%%%%%%%%%%%%%%%%%%%%%%%%%%%%%%%%%%%%%%%%%%%%%%%%%%%%%%%%%%%%%%%%%%%%%%%%%%%

\section{Quark Gap Equation: Truncation Schemes \label{trunc}}

In QCD, two-point Green functions, in particular the nonperturbative dressing of a current quark, are described by DSEs which are the relativistic equations 
of motion in that theory~\cite{Bashir:2012fs}. For arbitrary gauge and written in Euclidean space, the DSE of the  quark propagator is given by:
\begin{equation}
  S^{-1}_\xi(p)  = \, Z_2 \,  i\, \gamma\cdot p + Z_4 \, m(\mu)  
             + \, Z_1  4\pi \alpha_s^\xi \! \int^\Lambda\!\!  \frac{d^4k}{(2\pi)^4}\  \Delta^{ab}_{\mu\nu} (q)\, \gamma_\mu t^a\,  S_\xi (k) \,\Gamma_\nu^{b\xi} (k,p) \, .      
  \label{DSEquark}                             
\end{equation}
In this DSE,  $Z_1(\mu,\Lambda,\xi)$,  $Z_2(\mu,\Lambda,\xi)$ and $Z_4(\mu,\Lambda,\xi)$ are the vertex, wave function and mass renormalization constants, respectively, 
and $m(\mu)$ is the renormalized current-quark mass. In the self-energy integral, $\Lambda$ is a Poincar\'e-invariant cut-off, whereas  $\mu$ is the renormalization 
point chosen such that $\Lambda \gg \mu$. 

The quark-gluon interaction is described by the dressed vertex  $\Gamma^{a\xi}_\mu (k,p)=  \Gamma_\mu^\xi (k,p) t^a$, where $t^a = \lambda^a/2$  are the SU(3)
color-group generators in the fundamental representation, $p$ is the incoming and $k$ the outgoing quark momentum, respectively, while $q=k-p$ is the gluon momentum. 
The gluon propagator in $R_\xi$ gauge, 
\begin{equation}
    \Delta^{ab}_{\mu\nu} (q ) =  \delta^{a b} \left(\delta_{\mu \nu}-\frac{q_{\mu} q_{\nu}}{q^{2}}\right )  \Delta_\xi (q^2)  \,
       + \, \delta^{ab}\, \xi\dfrac{q_{\mu}q_{\nu}}{q^4} \ ,
 \label{gluonprop}   
\end{equation}    
is characterized by a transverse dressing function, renormalized as $\Delta_\xi(\mu^2 ) =1/\mu^2$. The gauge-parameter dependence of this  dressing function 
has been studied with lattice-QCD and functional approaches~\cite{Bicudo:2015rma,Huber:2015ria,Napetschnig:2021ria}.

The covariant decomposition of the DSE solutions is generally written in terms of two amplitudes, namely $A_\xi (p^2)$ and $B_\xi (p^2)$, 
\begin{equation}
\label{DEsol}
   S_\xi (p)  =    \frac{1}{i \gamma \cdot p \,A_\xi (p^2)  + B_\xi ( p^2 ) } \, = \,  \frac{Z_\xi (p^2 )}{ i \gamma \cdot p + M_\xi ( p^2 )} \, 
   = \,  -i \gamma \cdot p \, \sigma_{\rm v}^\xi ( p^2 ) + \sigma_{\rm s}^\xi ( p^2 )  \ ,
\end{equation}
where the  flavor- and gauge-dependent mass and wave renormalization functions can be expressed by the two momentum dependent amplitudes: 
\begin{equation}
      M_\xi (p^2) = B_\xi(p^{2},\mu^2 )/A_\xi(p^{2},\mu^2) \ ,  \qquad
     Z_\xi (p^2,\mu^2 ) = 1/A_\xi (p^2,\mu^2) \ .
\end{equation}
The renormalization scale we employ is low, $\mu = 4.3$~GeV, as it is the scale at which the transverse dressing function $\Delta_\xi (q^2)$ in $R_\xi$  
gauge is renormalized~\cite{Bicudo:2015rma} and allows for comparison of $M_\xi (p^2)$  and $Z_\xi (p^2,\Lambda^2 )$ with lattice QCD at this 
scale~\cite{Oliveira:2018lln}.  We therefore also impose the following conditions: $Z_\xi (\mu^2) =1$  and $M_\xi (\mu^2 ) \equiv  m(\mu^2) = 
25$~MeV~\cite{Rojas:2013tza,Rojas:2014tya,Serna:2018dwk}.

Coming back to the quark-gluon vertex introduced in Equation~\eqref{DSEquark}, its most general covariant  decomposition is not unique. For all that, one
can express this vertex most generally as~\cite{Ball:1980ay}, 
\begin{equation}
  \Gamma_{\mu}^\xi (k,p)  \, = \,  \Gamma^{L\xi}_{\mu}(k,p)  + \Gamma^{T\xi}_{\mu}(k,p) 
                                        \,  =  \, \sum_{i=1}^4  \, \lambda_i^\xi (k,p) L_{\mu}^i(k,p) + \sum_{i=1}^8 \tau_i^\xi (k,p) T_{\mu}^i(k,p)  \ ,
\label{BallChiu}
\end{equation}
where we stress the gauge dependence of the so-called longitudinal and transverse form factors, $\lambda_i^\xi (k,p)$ and $ \tau_i^\xi (k,p)$, respectively. 
The transverse vertex is naturally defined by  $i q \cdot \Gamma^{T\xi}  (k,p)  =  0$. The 12 vector structures $L_{\mu}^i(k,p) $ and $T_{\mu}^i(k,p)$ are build 
upon the three available vectors, namely $\gamma_\mu$, $k_\mu$ and $p_\mu$ and variations thereof, with the constraint that $\Gamma_{\mu}^\xi (k,p)$  
should exhibit the same transformation properties as the bare vertex $\gamma_\mu$ under charge conjugation $C$, parity transformation $P$ and time reversal $T$.

In Section~\ref{gaugedep}  we will observe the impact of a given truncation of Eq.~\eqref{DSEquark} on gauge covariance. To this end, we consider three cases, namely 
the leading rainbow truncation, a ghost-corrected Ball-Chiu vertex that satisfies the ``longitudinal'' STI~\cite{Rojas:2013tza,Aguilar:2010cn,Aguilar:2016lbe}, and the full 
longitudinal and transverse vertex structure derived with an additional transverse STI~\cite{Albino:2021rvj}. In the following, we present the form
factors $\lambda_i^\xi (k,p)$ and $ \tau_i^\xi (k,p)$ considered in each case, while the common basis elements $L_{\mu}^i(k,p) $ and $T_{\mu}^i(k,p)$ are listed
in the Appendix~\ref{appA}.    \medskip

%%%
\noindent
\emph{\underline{Rainbow truncation}}  \smallskip

$\Gamma^{\xi}_{\mu}(k,p) = \gamma_\mu\ $ :  the bare vertex, successfully employed with phenomenological interaction models to describe light hadrons~\cite{Chang:2009ae}
carries neither dynamical nor gauge information: $\lambda_1^\xi (k,p) = 1$, $\lambda_{2,3,4}^\xi (k,p) =  \tau_{1-8}^\xi (k,p) = 0$. This truncation is commonly employed with an 
interaction model or an artificially scaled-up strong coupling to compensate for the lack of support in the DSE kernel~\cite{Serna:2018dwk,El-Bennich:2016qmb}. Simple extensions 
include a flavor dependent coupling  in the treatment of heavy-light mesons and heavy quarkonia~\cite{Serna:2017nlr,Serna:2020txe,Serna:2022yfp,daSilveira:2022pte,Serna:2024vpn}. 
\bigskip

 %%% 
\noindent
\emph{\underline{STI Ball-Chiu vertex}}  \smallskip

$ \Gamma_{\mu}^\xi (k,p)  \, = \,  \sum_{i=1}^4   \lambda_i^\xi (k,p\,) L_{\mu}^i(k,p)$  :  this vertex is constructed to satisfy the STI, however the quark-ghost scattering
amplitude that enters this identity is modeled in a dressed-propagator approach and only the leading form factor $X_0^\xi (k,p)$ is kept; see Refs.~\cite{Rojas:2013tza,
Aguilar:2010cn,Aguilar:2016lbe,Davydychev:2000rt} for details. 
\begin{eqnarray}
  \lambda_1^\xi (k,p)  & = & \tfrac{1}{2}\,  G(q^2) X_0^\xi (q^2)\left[ A_\xi (k^2) + A_\xi (p^2)   \right] \ ,
 \label{lambda1QCD}   \\ [3pt]
   \lambda_2^\xi (k,p)  & = &  G(q^2) X_0^\xi (q^2)\, \frac{ A_\xi (k^2) - A_\xi(p^2)}{k^2-p^2}  \ ,
 \label{lambda2QCD}   \\ [3pt]
  \lambda_3^\xi (k,p)  & = &   G(q^2) X_0^\xi (q^2)\, \frac{ B_\xi (k^2) - B_\xi(p^2)}{k^2-p^2}  \ ,
 \label{lambda3QCD}   \\ [3pt]
  \lambda_4^\xi (k,p)  & = &\ 0 \ . 
 \label{lambda4QCD} 
\end{eqnarray} 
The dressing function $G(q^2)$ we use~\cite{Duarte:2016iko} is defined by the ghost propagator $D^{a b} (q^2) = -\, \delta^{ab} G(q^2)/ q^2$ 
and renormalized as $G(4.3^2\, \mathrm{GeV}^2) = 1$. Moreover, $\tau_1^\xi (k,p) = \tau_2^\xi (k,p)  = \ldots =   \tau_8^\xi (k,p)= 0$. 
\bigskip

\noindent
\emph{\underline{Full STI vertex}}  \smallskip

$\Gamma_{\mu}^\xi (k,p)  \, = \,  \sum_{i=1}^4  \, \lambda_i^\xi (k,p) L_{\mu}^i(k,p) + \sum_{i=1}^8 \tau_i^\xi (k,p) T_{\mu}^i(k,p)$  : the transverse STI constrains 
the vertex structures that are transverse to the gluon momentum and in addition to Equations~\eqref{lambda1QCD} to \eqref{lambda4QCD} one finds,  with the same approximation 
for the quark-ghost kernel~\cite{Albino:2021rvj}, the following transverse form factors: 
\begin{align}
   \tau_1^\xi (k,p) & =  - \frac{ Y_{1} }{ 2 (k^{2} - p^{2}) \nabla(k,p) } \ ,
\label{tau1QCD}   \\[3pt] 
   \tau_2^\xi (k,p) &= - \frac{Y_{5} - 3 Y_{3}}{ 4 (k^{2} - p^{2}) \nabla(k,p) } \ ,
\label{tau2QCD}  \\[3pt] 
   \tau_3^\xi (k,p) &=  \frac{1}{2}\,  G(q^2) X_0^\xi (q^2)\left[ \frac{ A_\xi (k^2) - A_\xi (p^2)}{k^2-p^2}  \right] +   \frac{Y_{2}}{4\nabla(k,p)}
   - \frac{ t^{2} (Y_{3} - Y_{5}) }{ 8(k^{2} - p^{2}) \nabla(k,p) } \ ,   
\label{tau3QCD}  \\[3pt] 
   \tau_4^\xi (k,p) &=  - \frac{ 6 Y_{4} + Y_{6}^{A} }{8\nabla(k,p) } - \frac{t^{2} Y_{7}^{S} }{ 8(k^{2} - p^{2}) \nabla(k,p) } \  ,
% \frac{Y_1\! - (k^2\! -p^2) ( 6 Y_4 + Y_6 )\! - (k+p)^{2} Y_7 }{ 2 (k^{2} - p^{2})^{2} \nabla(k,p) }  ,
\label{tau4QCD}  \\[3pt] 
   \tau_5^\xi (k,p) &= -  G(q^2) X_0^\xi (q^2)\left[ \frac{ B_\xi  (k^2) - B_\xi (p^2)}{k^2-p^2}  \right ]   - \frac{2 Y_4 + Y_6^A}{2 (k^{2}-p^{2})} \ ,
\label{tau5QCD}  \\[3pt] 
   \tau_6^\xi (k,p) &= \frac{(k-p)^{2} Y_{2} }{ 4 (k^{2} - p^{2}) \nabla(k,p) } -\frac{Y_3 - Y_5}{8\nabla(k,p) }  \ ,
\label{tau6QCD} % \\ [3pt] 
\end{align}
\begin{align}
   \tau_7^\xi (k,p) &= \frac{q^2 (6Y_4 +Y_6^A)}{4(k^2-p^2)\nabla(k,p)}   +\frac{Y_7^S}{4\nabla(k,p )} \ ,
\label{tau7QCD}  \\[3pt] 
  \tau_8^\xi (k,p) &= - G(q^2) X_0^\xi (q^2)\left[ \frac{ A_\xi  (k^2) - A_\xi (p^2)}{k^2-p^2}  \right] - \frac{2 Y_8^A}{k^{2}-p^{2}}  \ ,
%                     & - &  \frac{ Y_2(k-p)^2}{4 (k^2-p^2) \nabla(k,p) } - \frac{Y_3}{4\mabla (k,p)}  \, .
\label{tau8QCD}
\end{align}
where $Y_{i} \equiv Y_{i}^\xi (k,p)$ and the Gram determinant is defined by: $\nabla (k,p) = k^{2} p^{2} - (k \cdot p)^{2}$. In Equations~\eqref{tau1QCD} to \eqref{tau8QCD}, the
$Y_i^{(A,S)}$ functions are the form factors of the most general decomposition of a Fourier transformed four-point function in coordinate space. The latter involves a non-local 
vector vertex along with a Wilson line to preserve gauge invariance and contributes to the transverse STI. Since we deal with a rather complex object for which no calculations 
exist yet, we refer to the discussion in Ref.~\cite{He:2009sj} and note that the $Y_i^{(A,S)}$  were constrained~\cite{Albino:2018ncl} with a vertex ansatz guided 
by pQCD and multiplicative renormalizability~\cite{Bashir:2011dp}. We do stress the fact, however, that the functions $Y_1(k,p)$, $Y^A_6(k,p)$ and $Y^S_7 $ are 
\emph{massive}, i.e. they are proportional to the mass function $B(k^2)$. The contribution to DCSB of the vertex form factors in Equations~\eqref{lambda1QCD} to 
\eqref{tau8QCD} that depend directly on $B(k^2)$ or indirectly via these functions is therefore crucial and we observe that no mass-function solutions are found 
if all $Y_i(k, p)$ are zero, regardless of the value of the strong coupling $\alpha_s(\mu)$. The same occurs if one chooses $\tau_4(k, p)=\tau_7(k, p)=0$. 
Indeed, these two transverse form factors are responsible for the overwhelming contribution to DCSB in the gap equation~\cite{Albino:2021rvj}.

The gauge dependence of the strong coupling has been studied with the Schwinger mechanism in the three-gluon vertex, which is crucial to the generation of an effective 
gluon mass in the infrared domain~\cite{Aguilar:2016ock}:
\begin{equation}
  \alpha_s^\xi = \alpha_s^0 + 0.098 \xi - 0.064 \xi^2 \ .
\label{strongcoupling}  
\end{equation}
The coupling $\alpha_s^0 = 0.29$ is used for the ghost-corrected Ball-Chiu and the full STI vertices at $\mu=4.3$\;GeV. On the other hand, solving the DSE with this 
coupling in the rainbow truncation generates a very modest mass: $M_\xi^{u,d} (0) \approx 70$~MeV. In analogy with effective interaction models we therefore 
inflate the coupling strength to $\alpha_s^0 = 1.0$ and obtain a constituent mass similar to that found with the full STI vertex.

%%%%%%%%%%%%%%%%%%%%%%%%%%%%%%%%%%%%%%%%%%%%%%%%%%%%%%%%%%%%%%%%%%%%%%%%%%%%%%%%%%%%%%%%%
%%%%%%%%%%%%%%%%%%%%%%%%%%%%%%%%%%%%%%%%%%%%%%%%%%%%%%%%%%%%%%%%%%%%%%%%%%%%%%%%%%%%%%%%%

\section{Gauge Dependence of the Mass and Wave Renormalization \label{gaugedep} }

We now discuss the solutions, $M_\xi (p^2)$ and $Z_\xi (p^2)$, of Equation~\eqref{DSEquark} for the three cases considered in Section~\ref{trunc}. Henceforth, we make use 
of the gluon and ghost dressing functions from lattice QCD, $\Delta_\xi (q^2)$~\cite{Bicudo:2015rma} and $G(q^2)$~\cite{Duarte:2016iko} respectively, renormalized at 
$\mu =4.3$~GeV and parametrized  with a Pad\'e approximant~\cite{Dudal:2018cli},
\begin{equation}
   \Delta_\xi (q^2 )  =  Z \, \frac{q^{2}+M_{1}^{2}}{q^{4}+M_{2}^{2} q^{2}+M_{3}^{4}} \,
   \left [1 +  \omega \ln \left(\frac{q^{2}+M_0^2}{\Lambda_{\mathrm{QCD}}^2} \right ) \right ]^{\gamma_\mathrm{gl}}   ,
 \label{gluonparam}   
\end{equation}
where $\omega=11 N_c \,\alpha_{s}(\mu) /12 \pi$, $\Lambda_{\mathrm{QCD}}=0.425$~GeV and $\gamma_\mathrm{gl} =(13-3 \xi) / 22$ is the 1-loop anomalous 
gluon dimension. This parametrization is tantamount to a renormalization-group improved Pad\'e approximation and is motivated by the refined Gribov-Zwanziger 
tree-level gluon propagator in the infrared region. The parameters  $Z$, $M_0$, $M_1$, $M_2$ and $M_3$ were determined in a least-$\chi^2$ fit in 
Ref.~\cite{Lessa:2022wqc} and exhibit a \emph{nearly\/} linear dependence on the gauge parameter $\xi$ when fitted to the  lattice data for the six values, 
$\xi= 0, 0.1, 0.2, 0.3, 0.4, 0.5$.  This feature was used to extrapolate the gluon propagator up to Feynman gauge~\cite{Lessa:2022wqc}. 

The ghost propagator~\cite{Duarte:2016iko} is parametrized with a similar expression,
\begin{equation}
    G  (q^2)    = \,  Z\, \frac{q^4+M_2^2 q^2+M_1^4}{q^4+M_4^2 q^2+M_3^4} 
     \left [ 1 +\omega \ln \left(\frac{q^2+\frac{m_1^4}{q^2+m_0^2}}{\Lambda_\mathrm{QCD}^2} \right )  \right ]^{\gamma_{\mathrm{gh} } }   ,
  \label{ghostparam}   
\end{equation}
and is assumed to be independent of $\xi$~\cite{Cucchieri:2018leo}. The anomalous ghost dimension is given by $\gamma_{\mathrm{gh}} = -9/44$ while
$\omega$, $\Lambda_{\mathrm{QCD}}$ and $\mu$ are as in Equation~\eqref{gluonparam}. The fit parameters are also found in Ref.~\cite{Lessa:2022wqc}.
 
\begin{figure}[H]
%\vspace*{-7mm}
%\centering
\begin{adjustwidth}{-\extralength}{-2cm} 
\centering
  \includegraphics[scale=0.87]{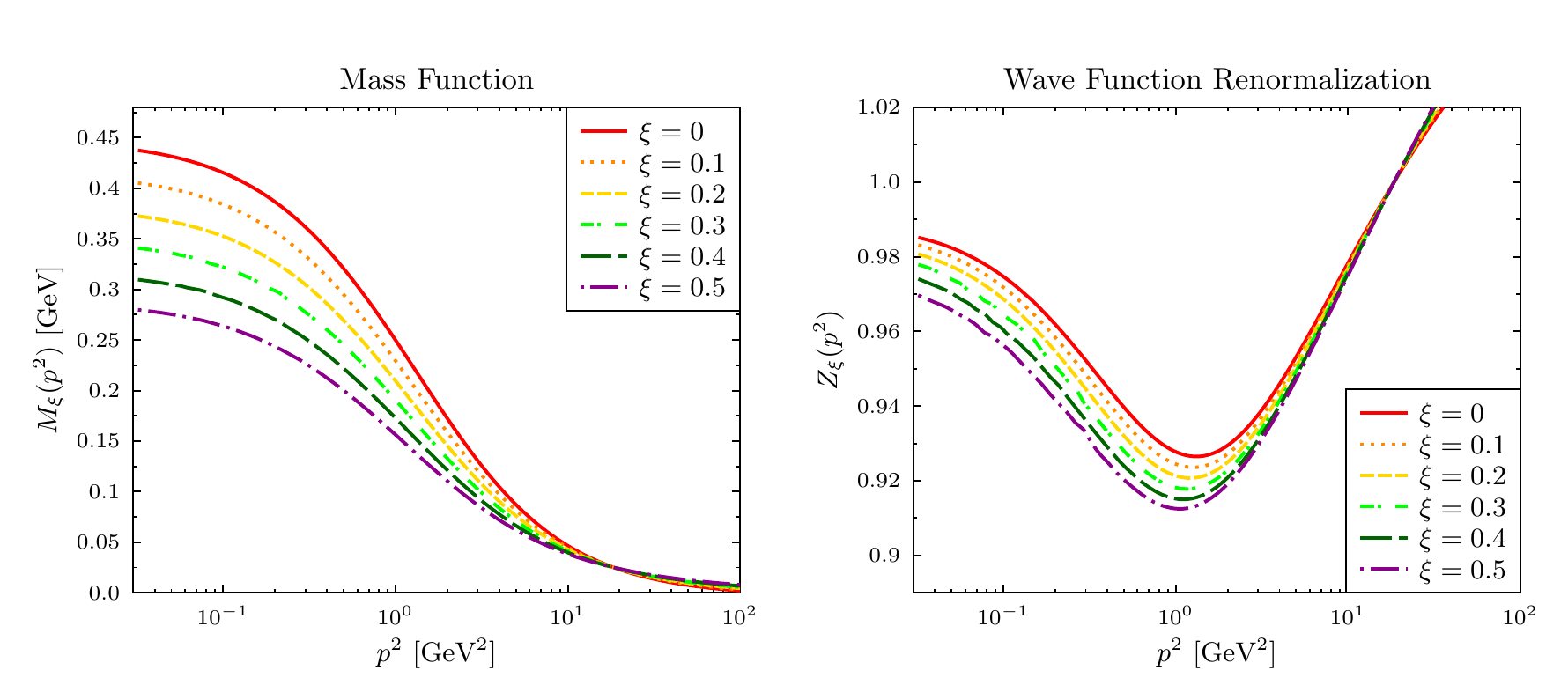} \end{adjustwidth} 
\caption{\label{figRL} Gauge-parameter
dependence of the $M_\xi (p^2)$ and $Z_\xi (p^2)$ functions of light quarks ($m_u = m_d$) in the rainbow truncation of the DSE.
Note that the strong coupling is large: $\alpha_s^0 =1.0$.}
\end{figure}

In Fig.~\ref{figRL} we plot the mass and wave-renormalization functions and their gauge-parameter dependence for the bare vertex. As mentioned earlier, with
$\alpha_s^0 \approx 0.3$ the dynamical mass generation is insufficient to produce a true constituent-quark mass for any gauge parameter. Hence, we inflate the constant
to $\alpha_s^0 =1.0$, as this will allow us to compare quark condensates and weak decay constants obtained with the full STI vertex later on. 
The gauge dependence of $M_\xi (p^2)$  is pronounced and the mass steadily decreases  as a function of the gauge parameter. This behavior was also observed 
in mass and wave-renormalization functions solving the DSE in quenched QED3 with  $\Gamma_\mu^\xi (k,p) = \gamma_\mu$~\cite{Bashir:2004yt}. Likewise, 
$Z_\xi (p^2)$ also decreases with $\xi$  in the range considered here.

%%%%%%%%%%%%%%%%%%%%%%%%%%%%%%%%%%%%%%%%%%%%%%%%%%%%%%%%%%%%%%%%%%%%%%%%%%%%%%%%%%%%%%%%%

\vspace{-18pt}
%%%%%%%%%%%%%%%%%%%%%%%%%%%%%%%%%%%%%%%%%%%%%%%%%%%%%%%%%%%%%%%%%%%%%%%%%%%%%%%%%%%%%%%%%

\begin{figure}[b!]
\vspace*{-2mm}
%\centering
\begin{adjustwidth}{-\extralength}{-2cm} 
\centering
  \includegraphics[scale=0.95]{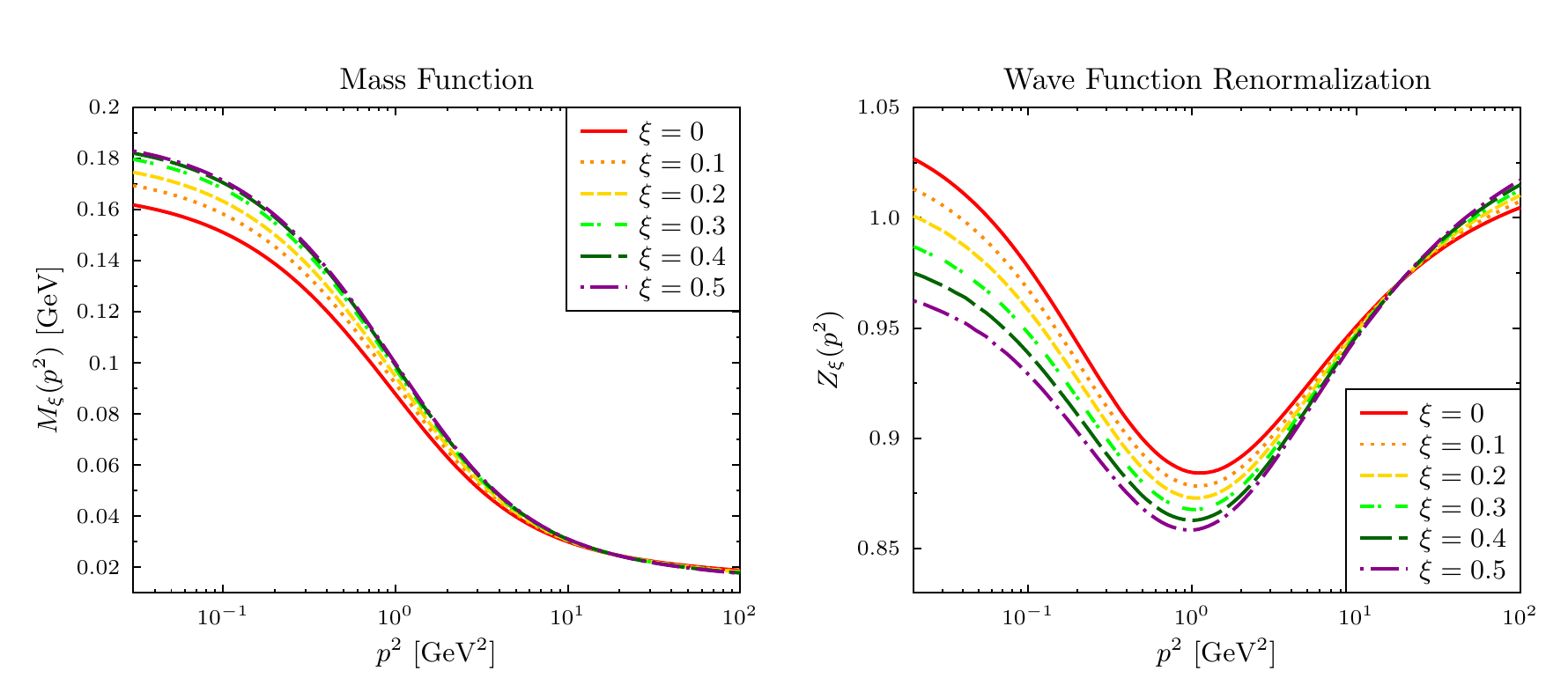}   \end{adjustwidth} 
   \caption{\label{figBC} The gauge-parameter dependence of $M_\xi (p^2)$ and $Z_\xi (p^2)$ of the light quark employing 
   the STI Ball--Chiu vertex, Equations~\eqref{lambda1QCD} to \eqref{lambda4QCD}, in the DSE~\eqref{DSEquark}. }
%
%\vspace*{-3mm}
\end{figure}

\begin{figure}[b!]
\begin{adjustwidth}{-\extralength}{-2cm} 
\centering
  \includegraphics[scale=0.95]{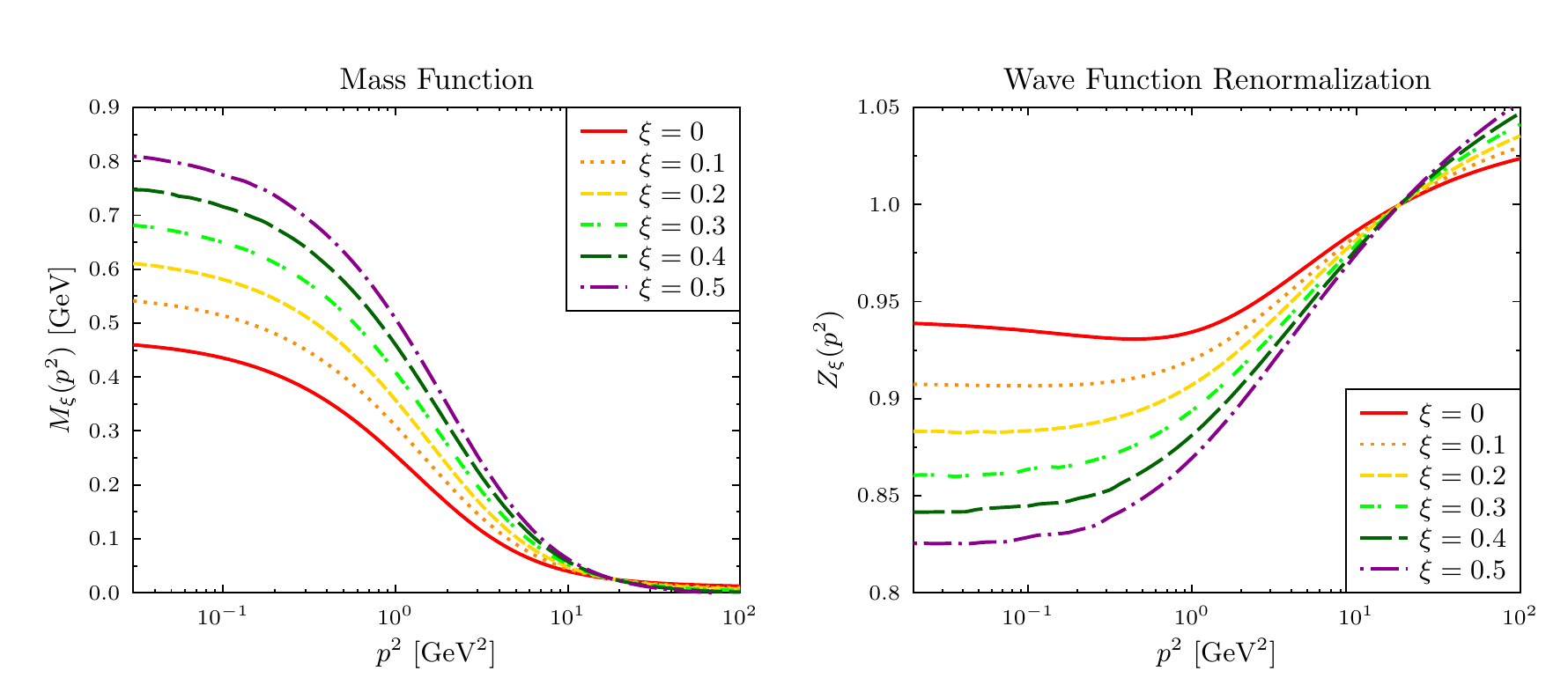}   \end{adjustwidth} 
  \caption{$M_\xi(p^2)$ and $Z_\xi(p^2)$ as functions of the gauge parameter $\xi$  DSE~\eqref{DSEquark} of the light quarks when the DSE~\eqref{DSEquark} is solved
               with the full STI vertex $\Gamma_\mu^\xi (k,p)$, Equations~\eqref{lambda1QCD} to \eqref{tau8QCD}. }
 \label{figMZ-Rxi} 
\end{figure}

Solving the DSE~\eqref{DSEquark} with the non-transverse vertex defined by Equations~\eqref{lambda1QCD} to \eqref{lambda4QCD}, we obtain the mass and wave-renormalization 
functions of Fig.~\ref{figBC}. We observe that the gauge dependence of the gap equation leads to a behavior of $M_\xi (p^2)$ opposed to that in Fig.~\ref{figRL}: the effect of 
the additional gauge dependence in the vertex is an increase of $M_\xi (p^2)$ up to $\xi=0.5$, at which point the function seemingly freezes. The functional behavior of 
$Z_\xi (p^2)$ exhibits the characteristic behavior of the Ball-Chiu vertex, with a local minimum at $p \approx 1\,$GeV and a sudden rise in the infrared, though the  
gauge dependence is more pronounced than in the rainbow truncation of Fig.~\ref{figRL}. We remark that the solutions of the mass and wave-renormalization functions
with the STI Ball-Chiu vertex are increasingly unstable for $\xi >0.5$. We therefore do not include the Feynman-gauge solutions in Fig.~\ref{figBC} and for comparison's 
sake we also refrain from doing so in Figs.~\ref{figRL} and \ref{figMZ-Rxi}.

%%%%%%%%%%%%%%%%%%%%%%%%%%%%%%%%%%%%%%%%%%%%%%%%%%%%%%%%%%%%%%%%%%%%%%%%%%%%%%%%%%%%%%%%%

%%%%%%%%%%%%%%%%%%%%%%%%%%%%%%%%%%%%%%%%%%%%%%%%%%%%%%%%%%%%%%%%%%%%%%%%%%%%%%%%%%%%%%%%%

In Fig.~\ref{figMZ-Rxi} we present our solutions for $M_\xi (p^2)$ and $Z_\xi (p^2)$ with the full gauge-dependent STI vertex defined by  Equations~\eqref{lambda1QCD} 
to \eqref{tau8QCD}. Clearly, the trend of a rising mass function and a dropping wave renormalization in the low-momentum region, observed with the ghost-corrected 
Ball-Chiu vertex, is corroborated when the transverse vertex is included. The contrast with the functional behavior in Fig.~\ref{figRL} is evident, in particular that of $Z_\xi (p^2)$ 
which now decreases steadily with $\xi$ and saturates in the infrared domain. Moreover, the quark-ghost scattering form factor,  $X_0^\xi (k,p)$, is calculated in a dressed approximation~\cite{Rojas:2013tza,Aguilar:2010cn,Aguilar:2016lbe} and dynamically generated in consistency with $M_\xi (p^2)$ and $Z_\xi (p^2)$. We abstain from 
discussing the gauge-dependent solutions here and refer to  Ref.~\cite{Lessa:2022wqc} for details. 

\begin{figure}[b!]
\centering
  \includegraphics[scale=0.8]{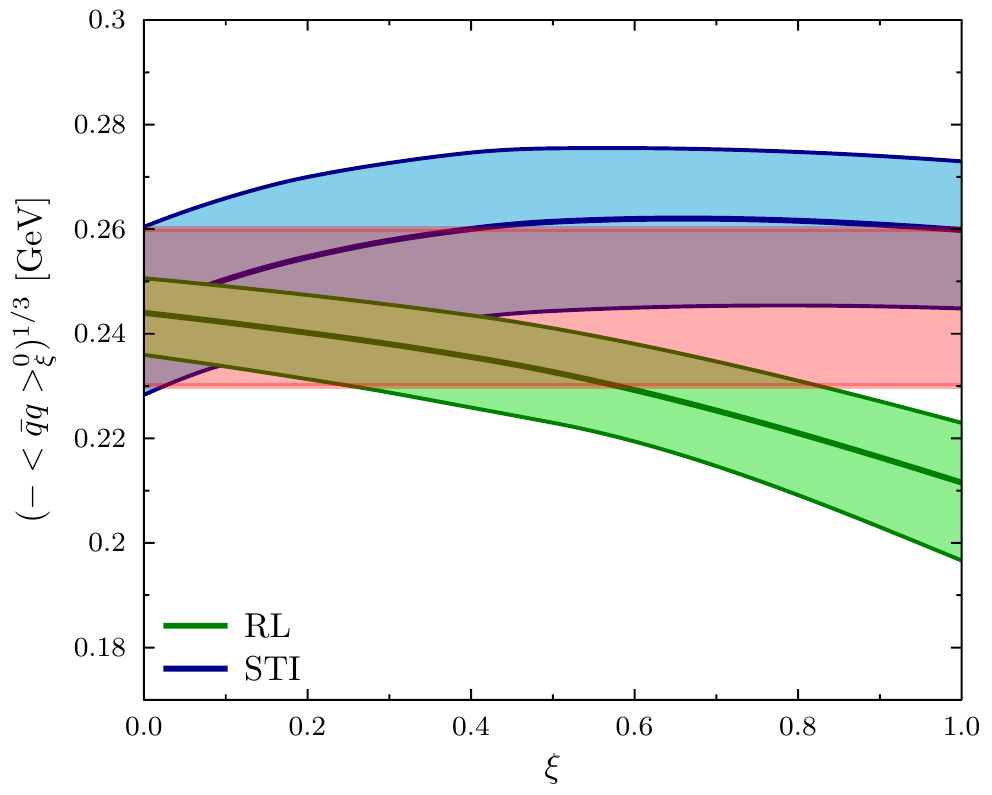}   
  \caption{Gauge dependence of the chiral quark condensate obtained with two different quark--gluon vertices in the DSE. 
  The pink-, green-, and blue-shaded error bands stem from the statistical errors of the lattice QCD predictions for the gluon and ghost propagators~\cite{Bicudo:2015rma} 
   %MDPI: Please confirm if copyright permission is required and should be obtained, and please add the corresponding copyright-related content in the caption if required (refer to https://www.mdpi.com/ethics#10)
   % There is no copyright. Lattice data are freely available and found in the reference mentioned. 
  on which the gauge-dependent solutions of $A_\xi(p^2)$ and  $B_\xi(p^2)$ depend via Equations~\eqref{DSEquark} and \eqref{gluonprop}.  
  The pink horizontal band indicates the region of a gauge-independent quark condensate as implied by the LKFTs in QCD, 
 where the uncertainty is due to the statistical error of the  gauge propagators in Landau gauge.}
    \label{quarkcond}
\end{figure}

To conclude our comparison of gauge dependent quark propagators, we ought to ask which is the correct one. This question can only be answered by computing a 
gauge-invariant quantity with the given quark propagators. A simple quantity that does not require the solutions of a Bethe-Salpeter equation is the quark condensate
in the chiral limit ($\Lambda =1$~GeV), 
\begin{equation}
   -  \langle\bar{q} q \rangle^0_\xi  \, \equiv \, Z_4  N_c  \int^{\Lambda}  \!  \frac{d^4k}{(2\pi)^4} \, \operatorname{tr}_{D}  \left [ S_\xi^0 (k) \right ] \ ,
\end{equation}
which we can express as a function of $\xi$. As can be inferred from Fig.~\ref{quarkcond},  the condensate calculated with the full STI vertex is the central line in 
the blue-shaded, fairly horizontal band. The latter depicts our error estimate due to the statistical error of the gluon propagator~\cite{Bicudo:2015rma}. A systematic 
error is not included, but we allow for uncertainties of the ghost dressing function and of $\alpha_s^\xi$~\eqref{strongcoupling} by varying $\Delta_\xi (q^2)$ by 
$\pm 5$\%. In this light, one can argue that the QCD condensate exhibits a moderate dependence on $\xi$ of the order of 7--8\%, rising initially from Landau 
gauge to $\xi \approx 0.3$ where the condensate seems to level off. This is in accordance with an invariance proof for any SU($N$) theory derived in 
Ref.~\cite{Aslam:2015nia} and based on the LKFT. The same figure also demonstrates that this is not the case for a quark condensate computed in the rainbow 
truncation of the DSE. In fact, while an extrapolation beyond $\xi =1.0$ of the gluon propagator must be taken with a grain of salt, we verified that the condensate 
in this simplest truncation keeps falling off with the gauge parameter. This functional behavior expresses the lack of gauge covariance in this leading truncation 
not constrained by the STI and multiplicative renormalizability.

%%%%%%%%%%%%%%%%%%%%%%%%%%%%%%%%%%%%%%%%%%%%%%%%%%%%%%%%%%%%%%%%%%%%%%%%%%%%%%%%%%%%%%%%% 

%%%%%%%%%%%%%%%%%%%%%%%%%%%%%%%%%%%%%%%%%%%%%%%%%%%%%%%%%%%%%%%%%%%%%%%%%%%%%%%%%%%%%%%%%

We can also calculate the weak decay constant of the pion in the chiral limit following Ref.~\cite{Roberts:1994hh}, as $f_\pi$ does not vanish in this limit whereas
the pion's mass does in truncations that preserve the axialvector WGTI. The weak decay constant in the chiral limit can be expressed by the integral,
\begin{align}
  \left  ( f_{\pi}^0 \right )^{\! 2} \,  = \,  \dfrac{N_c}{8 \pi^2}  \int_0^\infty \! dp^2 p^2  &\, B^2(p^2)  \left  (
                           \sigma_{\rm v}^2 - 2  \left [ \sigma_{\rm s}\sigma_{\rm s}^{\prime} +p^2 \sigma_{\rm v} \sigma_{\rm v}^{\prime} \right ]  \right.
 \nonumber \\
                     & \left.  - \ p^2 \left [ \sigma_{\rm s}\sigma_{\rm s}^{\prime \prime} - (\sigma_{\rm s}^{\prime} )^2 \right  ]
                                  - p^4 \left [ \sigma_{\rm v} \sigma_{\rm v}^{\prime \prime} - (\sigma_{\rm v}^{\prime} )^2 \right ]  \right ) \ ,
\label{fpi}
\end{align}
with $\sigma_{\rm s,v}^\prime \equiv d\sigma_{\rm s,v} (p^2)/dp^2$ and where we dropped the gauge-dependence label. As we are constrained by a low renormalization 
point due to the quenched lattice-QCD input for the gluon and ghost propagators we employ, we set $m(\mu)= 0$~MeV at $\mu = 4.3$~GeV, which allows for a sensible 
approximation for $\sigma_{\rm s}(p^2)$ and  $\sigma_{\rm v} (p^2)$ in Equation~\eqref{fpi}.

\begin{figure}[t!]
\centering
  \includegraphics[scale=0.8]{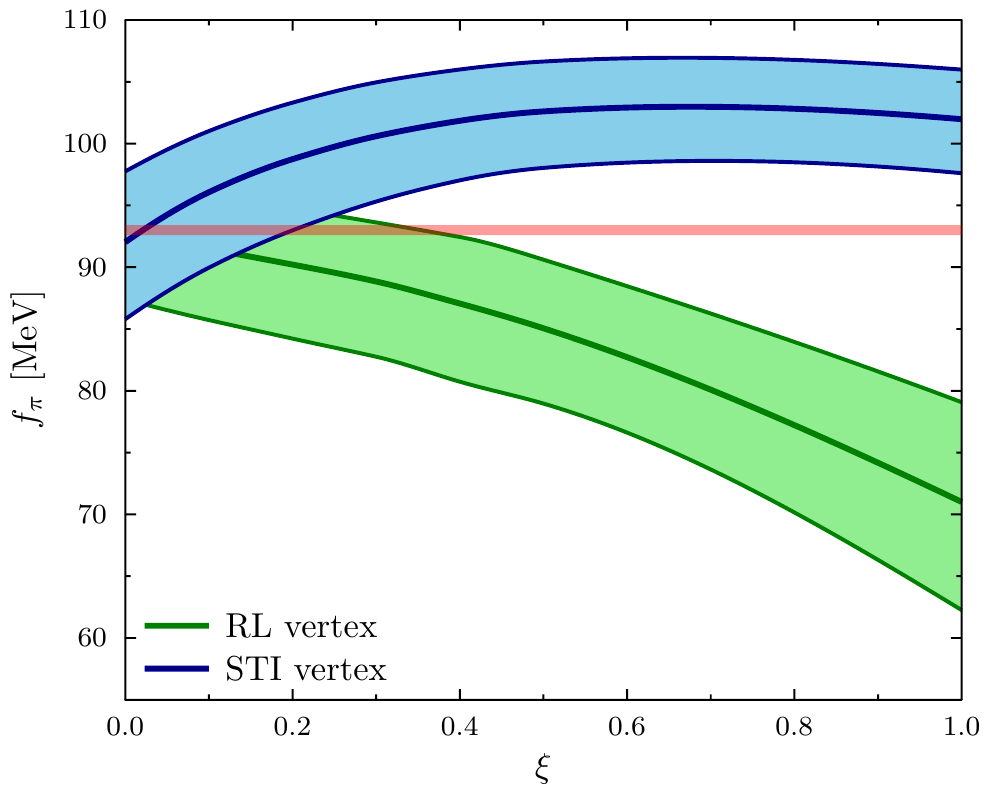}   
  \caption{Gauge dependence of the pion decay constant for two different quark--gluon vertices in the DSE~\eqref{DSEquark}. The error estimates
  represented by the green and blue bands are as in Figure~\ref{quarkcond}, while the horizontal pink line represents the experimental reference value.}
  \label{STI-RL-fpi}
\end{figure}

In analogy with the condensates presented in Fig.~\ref{quarkcond}, the rainbow truncation of the DSE yields a weak decay constant that decreases relatively fast, 
as can be appreciated in  Fig.~\ref{STI-RL-fpi}. If we solve the DSE with the full STI vertex, we observe an initial increase of $f_\pi$ from $\xi=0$ to $\xi \approx 0.5$
after which the decay constant appears to saturate. The variation between Landau and Feynman gauge is about 10\%, larger than the 7\% increase of the quark
condensate as function of $\xi$. However, this moderate effect is not yet conclusive, as we ought to compute $f_\pi$ with the complete Bethe-Salpeter 
amplitude of the pion in this much more challenging truncation.

%%%%%%%%%%%%%%%%%%%%%%%%%%%%%%%%%%%%%%%%%%%%%%%%%%%%%%%%%%%%%%%%%%%%%%%%%%%%%%%%%%%%%%%%% 

%%%%%%%%%%%%%%%%%%%%%%%%%%%%%%%%%%%%%%%%%%%%%%%%%%%%%%%%%%%%%%%%%%%%%%%%%%%%%%%%%%%%%%%%%
%%%%%%%%%%%%%%%%%%%%%%%%%%%%%%%%%%%%%%%%%%%%%%%%%%%%%%%%%%%%%%%%%%%%%%%%%%%%%%%%%%%%%%%%%

\section{Final Remarks}

We have shown that within a functional approach to QCD the transverse quark-gluon vertex plays an eminent role in the dynamics of mass generation and the emergence 
of a constituent quark mass scale.  Regarding the gauge parameter dependence of the gap equation, we find a running mass function that increases with $\xi$ and a 
quark wave function which  is mostly  sensible to the gauge parameter in the infrared domain. More precisely, $Z(0)$ drops from about 0.94 in Landau gauge to 0.83 
for $\xi =0.5$, a $7\%$ effect. As far as gauge invariance is concerned, despite the limitations of the setup used herein, a variation of about 7\% is observed in the 
quark condensate when the full vertex is taken into account in the gap equation. This contrasts with the outcome of the rainbow approximation which yields a condensate 
that decreases with $\xi$. Likewise, $f_\pi$ rapidly falls off as a function of the gauge parameter. These results illustrate, once more, the important role of the transverse 
vertex in the nonperturbative dynamics of the gap equation.

%%%%%%%%%%%%%%%%%%%%%%%%%%%%%%%%%%%%%%%%%%%%%%%%%%%%%%%%%%%%%%%%%%%%%%%%%%%%%%%%%%%%%%%%%
%%%%%%%%%%%%%%%%%%%%%%%%%%%%%%%%%%%%%%%%%%%%%%%%%%%%%%%%%%%%%%%%%%%%%%%%%%%%%%%%%%%%%%%%%

% The order of the section titles is different for some journals. Please refer to the "Instructions for Authors” on the journal homepage.
%%%%%%%%%%%%%%%%%%%%%%%%%%%%%%%%%%%%%%%%%%
\vspace{6pt}

\funding{This research was funded  by the S\~ao Paulo Research Foundation (FAPESP), grant no.~2023/00195-8  and by the National Council for Scientific and Technological 
 Development (CNPq), grant no.~409032/2023-9.  B.\,E. is a member of the Brazilian nuclear physics network project \emph{INCT-F\'isica Nuclear e Aplica\c{c}\~oes\/}, 
 grant no.~464898/2014-5.}

%\dataavailability{\hl{Data is contained within the article.} %MDPI: We encourage all authors of articles published in MDPI journals to share their research data. In this section, please provide details regarding where data supporting reported results can be found, including links to publicly archived datasets analyzed or generated during the study. Where no new data were created, or where data is unavailable due to privacy or ethical restrictions, a statement is still required. Suggested Data Availability Statements are available in section ``MDPI Research Data Policies'' at \url{https://www.mdpi.com/ethics}
%}

% REPLY: There is no "data" in this paper, the results can be reproduced with the analytical expressions given in this work using any kind of numerical method to solve the integrals. 

\acknowledgments{Much of what I presented in this contribution has only been possible with the precious input and contributions by Luis Albino, Adnan Bashir, 
Roberto Correa da Silveira, Jos\'e Lessa, Orlando Oliveira and Fernando Serna over the past years. I would like to express my gratitude to Kazuo Tsushima,
Anthony Thomas and Myung Ki Cheoun for the opportunity to contribute to the special  Symmetry issue \emph{Chiral Symmetry, and Restoration in Nuclear Dense Matter}. }

\conflictsofinterest{The author declares no conflicts of interest.} 

%%%%%%%%%%%%%%%%%%%%%%%%%%%%%%%%%%%%%%%%%%
%% Optional

%% Only for journal Encyclopedia
%\entrylink{The Link to this entry published on the encyclopedia platform.}

\abbreviations{Abbreviations}{
The following abbreviations are used in this manuscript:\\

\noindent 
\begin{tabular}{@{}ll}
 QCD & Quantum chromodynamics\\
 QED & Quantum electrodynamics \\
 DSE & Dyson--Schwinger equation \\
 LFKT &  Landau--Khalatnikov--Fradkin transformations \\
 WGTI & Ward--Green--Takahashi identity  \\
 STI  & Slavnov--Taylor identity
\end{tabular}
}

%%%%%%%%%%%%%%%%%%%%%%%%%%%%%%%%%%%%%%%%%%
%% Optional
\appendixtitles{no} % Leave argument "no" if all appendix headings stay EMPTY (then no dot is printed after "Appendix A"). If the appendix sections contain a heading then change the argument to "yes".

\appendixstart
\appendix
\section[\appendixname~\thesection]{}\label{appA}

For the longitudinal vertex we employ the common Ball-Chiu decomposition,
\begin{eqnarray}
L^{1}_{\mu}(k,p) &=& \gamma_\mu   \ ,   \\   [3pt] 
L^{2}_{\mu}(k,p) &=&  \tfrac{1}{2}\,  t_\mu\,  \gamma \cdot t \, , \\ [3pt] 
L^{3}_{\mu}(k,p) &=&     -  i\, t_\mu  \,,  \\ [3pt] 
L^{4}_{\mu}(k,p) &=&  -  \sigma_{\mu\nu} t_\nu    \, ,
\end{eqnarray}
where $t=k+p$ and $\sigma_{\mu \, \nu} = \frac{i}{2} \left[ \gamma_{\mu},\gamma_{\nu} \right ]$ in Euclidean space. 
The transverse vertex can generally be decomposed into eight independent vector basis elements. For the kinematical configuration discussed below 
Equation~\eqref{DSEquark}, they are given by,

\begin{eqnarray}
T^{1}_{\mu}(k,p) &=& i \!\left[ p_{\mu} (k \cdot q) -k_{\mu} (p \cdot q) \right] \,,    \\   [3pt] 
T^{2}_{\mu}(k,p) &=& \left [ p_{\mu} (k \cdot q) -k_{\mu} (p \cdot q) \right] \gamma \cdot t \,, \\ [3pt] 
T^{3}_{\mu}(k,p) &=& q^{2} \gamma_{\mu} - q_{\mu}\, \gamma \cdot q \,,  \\ [3pt] 
T^{4}_{\mu}(k,p) &=&  i q^{2}\left[\gamma_{\mu} \gamma \cdot t-t_{\mu}\right]+2 q_{\mu} p_{\nu} k_{\rho} \sigma_{\nu \rho} \, , \\ [3pt] 
T^{5}_{\mu}(k,p) &=& \sigma_{\mu \, \nu} q_{\nu}  \,,  \\  [3pt] 
T^{6}_{\mu}(k,p) &=& -\, \gamma_{\mu} \left( k^{2}-p^{2} \right) + t_{\mu}\, \gamma \cdot q \,,  \\  [3pt] 
T^{7}_{\mu}(k,p) &=&  \tfrac{i}{2} (k^{2}-p^{2}) \left[ \gamma_{\mu} \gamma \cdot t - t_{\mu} \right] + t_{\mu}\, p_{\nu} k_{\rho} \sigma_{\nu  \rho}   \, ,  \\ [3pt] 
T^{8}_{\mu}(k,p) &=& - \, i \gamma_{\mu} \, p_{\nu} k_{\rho} \sigma_{\nu  \rho} - p_{\mu} \, \gamma \cdot k + k_{\mu}\, \gamma \cdot p \, ,
\end{eqnarray}
We use a slightly modified basis~\cite{Kizilersu:1995iz} with respect to the Ball-Chiu vertex~\cite{Ball:1980ay} with the effect that all
transverse form factors are independent of kinematic singularities in one-loop perturbation theory and arbitrary covariant gauge.

%%%%%%%%%%%%%%%%%%%%%%%%%%%%%%%%%%%%%%%%%%%%%%%%%%%%%%%%%%%%%%%%%%%%%%%%%%%%%%%%%%%%%%%%%
%%%%%%%%%%%%%%%%%%%%%%%%%%%%%%%%%%%%%%%%%%%%%%%%%%%%%%%%%%%%%%%%%%%%%%%%%%%%%%%%%%%%%%%%%

%%%%%%%%%%%%%%%%%%%%%%%%%%%%%%%%%%%%%%%%%%
\begin{adjustwidth}{-\extralength}{0cm}
%\printendnotes[custom] % Un-comment to print a list of endnotes

\reftitle{References}

\PublishersNote{}
\end{adjustwidth}
\end{document}